\documentclass[prl,aps,twocolumn,showpacs]{revtex4}
\usepackage{graphicx}
\begin{document}

\title{Space Charge Limited Current Revisited: the Effect of \\ Surface Traps.}
\author{R. W. I. de Boer} \author{A. F. Morpurgo}
\affiliation{Kavli Institute of Nanoscience, Faculty of Applied
Sciences, Delft University of Technology, Lorentzweg 1, 2628 CJ
Delft, the Netherlands}
\date{\today}

\begin{abstract}

We analyze the effect of surface traps on unipolar space charge
limited current and find that they have a profound influence on
the $I-V$ curves. By performing calculations that account for the
presence of these traps, we can reproduce experimental
observations not captured by the conventional theory that only
considers the presence of traps in the bulk of the material.
Through the use of realistic material parameters, we show that the
effects discussed have clear experimental relevance.

\end{abstract}

\pacs{72.20.Ht, 72.80.Le, 73.20.At}

\maketitle

Space charge limited transport occurs in un-doped, wide-gap
semiconductors in which the density of charge carriers at
equilibrium is vanishingly small. In these materials, the current
is carried by charge injected from the contacts, whose density is
determined (limited) by electrostatics. Space charge limited
current (SCLC) is relevant for the operation of electronic devices
(e.g., organic light emitting diodes) and is routinely used in the
characterization of novel semiconductors to estimate parameters
such as the mobility of charge carriers, the density of trap
states, and their energy depth
\cite{Campbell97,Podzorov04,DeBoer04,Lang04,Jurchescu04}.

The description of SCLC relies on a simple phenomenological theory
first developed in the fifties and extended later to include model
specific details \cite{Lampert70,Kao04}. These extensions have led
to predictions of SCLC $I-V$ characteristics that resemble rather
closely what is actually measured. In practice, however, the
comparison between space charge limited current (SCLC)measurements
and theory is not very satisfactory in many cases, especially for
experiments performed on high-quality materials with a low density
of defects. Firstly, an independent validation of specific
assumptions adopted in the analysis of SCLC $I-V$ curves is almost
always impossible, which causes ambiguities in the interpretation
of the experiments. Even when the theoretically predicted $I-V$
characteristics exhibit a behavior close to that observed
experimentally, a sufficiently detailed analysis often reveals
inconsistencies \cite{Pope99,Campos72}. Secondly, SCLC $I-V$
curves measured on nominally identical samples often exhibit
significant differences \cite{DeBoer04}, which makes it
inappropriate to compare experimental data with theoretical
predictions that are critically sensitive to the assumptions on
which models rely. The current situation suggests that some
important aspects of the physics of space charge limited transport
are being overlooked.

In this context, and motivated by recent experimental work on
high-quality organic single crystals, we investigate the effect of
deep traps at the surface of the semiconducting material,
underneath the electrical contacts used to inject charge carriers.
We show that these surface traps are an essential ingredient for
the proper understanding of SCLC $I-V$ curves, which has been
neglected until now. Using a simple extension of the original
theory, we analyze the effect of these traps and demonstrate that
they cause a large change in the electrostatic profile in the bulk
of the material, thereby profoundly affecting the behavior of
SCLC. Calculations accounting for the presence of surface traps
enable us to reproduce experimental observations that are not
captured by the conventional theory, such as orders-of-magnitude
asymmetries in the $I-V$ curves. These calculations further
illustrate how the combined effect of surface and bulk traps also
results in features in the $I-V$ curves that have been so far
attributed to different physical mechanisms, which may explain the
inconsistencies found in the analysis of past experimental
results.

To understand how the effect of surface traps is taken into
account in our calculations, we first briefly review the main
aspects of the conventional theory of unipolar SCLC
\cite{Lampert70}. The theory relies on the simultaneous solution
of the Poisson and the continuity equation,
\begin{equation} \label{Poisson}
\frac{dE(x)}{dx} = \frac{e n_s(x)}{\epsilon}
\end{equation}
and
\begin{equation} \label{continuity}
J(x)= e n_f(x) \mu E(x) = \mathrm{constant},
\end{equation}
that relate the electric field $E$ and the current density $J$
($x$ is the distance from the injecting electrode). Here $n_s(x)$
is the total space charge at position $x$ and $n_f(x)$ is the part
of space charge that is free to move, which is smaller than
$n_s(x)$ if deep traps are present. Given the bulk density of
traps $N_t$ and their energy depth $E_t$, a relation between
$n_f(x)$ and $n_s(x)$ can be found via the Fermi-Dirac
distribution. This enables the solution of the equations above
that gives the relation between $J(x)$ and $E(x)$, i.e., the $I-V$
characteristics of the material.

The bulk density of deep traps $N_t$ is usually assumed to be
uniform throughout the material. In real materials, however, many
more traps per unit volume are likely to be present close the
sample surface than in the bulk. These surface traps can have
different physical origins. In inorganic covalently bonded
materials, for instance, they may be due to dangling bonds
resulting in the presence of surface states whose energy is deep
inside the semiconducting gap. In organic systems, such as van der
Waals bonded molecular single crystals, traps can originate from
molecules at the surface that are damaged during the contact
fabrication process. For pure semiconducting materials in which
the bulk density of traps is low, the \textit{total number of
traps at the sample surface can be comparable or even larger than
the total amount of bulk traps} even for rather thick samples.
This corresponds to a physical regime whose relevance has not been
sufficiently appreciated in the past. In particular, the effect of
trapped surface charge on the electrostatic profile throughout the
entire bulk of the sample has never been analyzed theoretically.

The modification of the conventional SCLC theory needed to take
into account the electrostatic effects due to surface traps is
minimal. It is sufficient to allow $N_t$ to depend on position, so
that its value in the region close to the contacts is much larger
than the bulk value. The analysis of this case is carried out by
solving numerically Eq. \ref{Poisson} and \ref{continuity}, from
which we directly obtain the $I-V$ characteristics.

\begin{figure}[t]
\centering
\includegraphics[width=7.5cm]{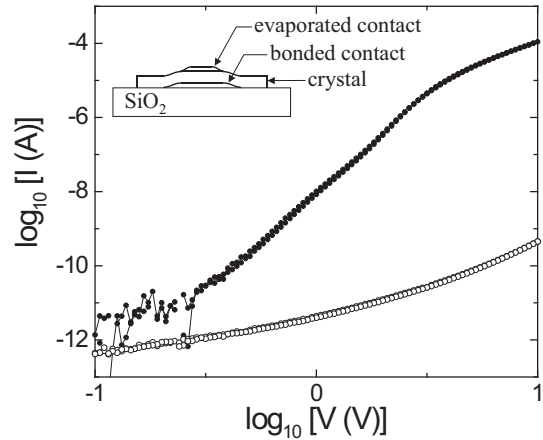}
\caption{$I-V$ characteristics of a 1-$\mu$m-thick tetracene
single crystal, measured for opposite polarities of the bias
voltage. The full symbols correspond to the current measured when
holes are injected from the bottom (electrostatically bonded)
contact; the open symbols correspond to the case of holes injected
from the top (evaporated) contact. The device configuration is
shown in the inset. \label{asymmetry}}
\end{figure}

The most interesting case is that in which the density of surface
traps present under one of the contacts is considerably larger
than that present under the other contact. This is experimentally
relevant because in general bottom and top contacts (see inset
Fig.~\ref{asymmetry}) are prepared in different ways, which
results in different surface trap densities. Also for contacts
prepared in a nominally identical way, the surface density of
traps at the two contacts can differ considerably because the
defects introduced at the surface depend on unknown parameters
that are not under experimental control
\cite{DeBoer04,Reynaert05}. We compare the simplest possible
situations in which surface traps are present only under one of
the two contacts (bulk traps are also present): the case of traps
present at both contacts with different densities can be analyzed
in an identical way and does not add new physics.

Our considerations are valid in different classes of materials.
Nevertheless, here we will mainly have in mind the case of organic
molecular single crystals (such as tetracene, rubrene, the metal
phthalocyanines, etc.), to which we have recently devoted
considerable experimental effort. This enables us to insert in our
calculations realistic values of the parameters and
to prove that the effect of surface traps is relevant in actual
experiments.

An illustrative example of SCLC through tetracene (a hole
conductor) crystals is shown in Fig.~\ref{asymmetry}. The data are
measured on an approximately 1-$\mu$m-thick single-crystal
contacted with two gold electrodes prepared in different ways. One
of the contacts is fabricated by placing the thin crystal onto a
metal film deposited on a substrate to which the thin, flexible
crystal adheres spontaneously. This procedure is known to result
in high-quality electrical contacts between the metal and the
crystal. The other contact is prepared by electron-beam
evaporation of gold onto the crystal, which is known to cause
"damage" to the crystal due to the exposure of the crystal surface
to X-Ray and high-energy electrons generated by the electron-beam,
during the evaporation process. The most striking feature of the
SCLC $I-V$ curves is the large (five to six orders of magnitude)
asymmetry: the measured current depends very strongly on the
contact used to inject holes into the material. Order-of-magnitude
asymmetries are regularly found for different contact fabrication
techniques.

\begin{figure}[t]
\centering
\includegraphics[width=7.5cm]{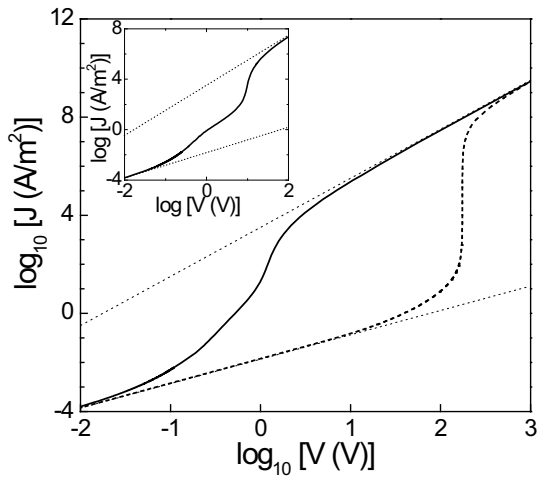}
\caption{Calculated SCLC $I-V$ curve for two different polarities
of the applied bias. The continuous (dashed) line represents the
current flowing when holes are injected from the contact without
(with) surface traps. Note the large asymmetry, which reproduces
the experimentally observed behavior (see Fig.~\ref{asymmetry}).
The inset shows that, when the surface traps are located at the
injecting contact, the amount of surface and bulk traps can be
chosen so that the transition to the trap-filled limit exhibits
multiple steps. \label{simulationasymmetry}}
\end{figure}

Fig.~\ref{simulationasymmetry} shows the $I-V$ characteristics
calculated for a sample with bulk traps and surface traps under
only one of the two contacts. The separation between the contacts
is taken to be 1~$\mu$m, corresponding to the case of the device
whose data are shown in Fig.~\ref{asymmetry}. The bulk density of
traps is taken to be $1 \cdot 10^{14}$ cm$^{-3}$. This is a
conservative, realistic estimate \cite{DeBoer04} and lower values
\cite{Jurchescu04} would result in a more pronounced effect of the
surface traps. The surface density of traps is set to correspond
approximately to one trap per every 1000 molecules in the first
few molecular layers of the crystal, which we model as a region of
approximately 10 nm at the crystal surface, containing a density
of traps of $\sim 3 \cdot 10^{18}$ cm$^{-3}$. For both bulk and
surface traps, the energy depth is taken to be same (to avoid the
insertion of unnecessary additional parameters) and equal to $0.7$
eV, with the precise value of this parameter not being critical
for our conclusions. Other parameters required by the model (but
which do not have any considerable influence on the results) are
the hole mobility, which only contributes as an overall scale
factor (we take $\mu = 1$ cm$^2$/Vs), the density of dopants and
their energy depth \cite{parameters}.

The calculations show that in the range between 1 and 100 V the
current depends very strongly on the voltage polarity. This
reproduces the huge asymmetry in the $I-V$ characteristics
observed experimentally. In agreement with the experiments, the
current is large when the polarity is such that holes are injected
from the contact free of surface traps. This result, which is
robust and does not critically depend on the values of the
parameters chosen above, clearly demonstrates the relevance of
surface traps.

\begin{figure}[t]
\centering
\includegraphics[width=8cm]{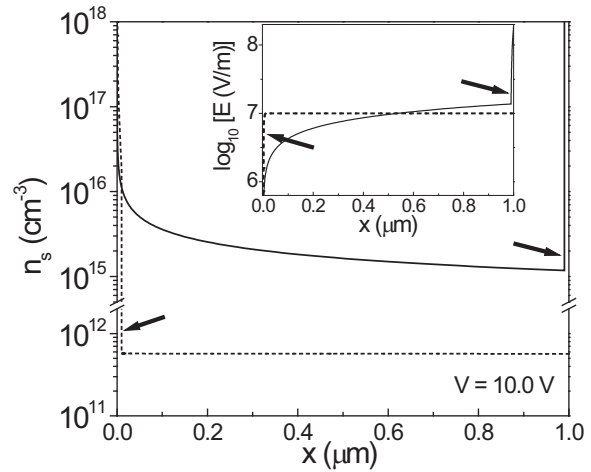}
\caption{Profile of the total (trapped plus free) density of
charge in the sample whose $I-V$ curves are shown in
Fig.~\ref{asymmetry}, for a 10 V applied bias. The injecting and
extracting contacts are located at $x=0$ and $x=1 \ \mu$m,
respectively. The continuous (dashed) line corresponds to the case
where the surface traps (pointed to by the arrows) are at the
extracting (injecting) contact. A large density of charge is
injected into the bulk when the surface traps are located at the
extracting contact, but not when the traps are located at the
injecting contact. In the inset, the electric field profile is
shown in the two cases. \label{spacecharge}}
\end{figure}

Microscopically, the asymmetry originates from a large difference
in the electrostatic profile in the bulk of the samples, for the
two bias polarities. This is illustrated in
Fig.~\ref{spacecharge}, where the density of charge is plotted as
a function of position for the two polarities, with 10~V applied
bias. When traps are located at the injecting contact, the bias is
not sufficient to reach complete filling of the surface traps, so
that no charge is injected in the bulk. The current is then
carried only by thermally activated carriers and its magnitude is
therefore small. On the contrary, the same bias is largely
sufficient to fill the same amount of traps at the extracting
contact. In this case the injected charge is present throughout
the bulk of the sample, enabling a large current flow. We conclude
that surface traps at the injecting contact suppress charge
injection and the current flow much more drastically than trap at
the extracting contact. As we will discuss at the end, this
behavior can be easily understood qualitatively.

Fig.~\ref{simulationasymmetry} also shows that the \textit{shape}
of the $I-V$ curve differs, depending on the contact used to
inject the carriers. If the carriers are injected from the
contacts where the surface traps are, the transition is very sharp
and similar to that predicted by the conventional theory when only
bulk traps at a discrete energy value are present. However, if the
carriers are injected from the trap-free contact, the transition
from the linear to the trap-filled regime (where $I \propto V^2$)
is smooth and power-law like, with an exponent larger than~2. In
this case, the precise shape of the transition is governed by the
ratio between the density of surface traps at the extracting
contact and the density of bulk traps. This is illustrated in the
inset of Fig.~\ref{simulationasymmetry}, where this ratio has been
changed to produce two apparent transitions in the $I-V$ curve. We
note that in the analysis of experimental SCLC curves based on the
conventional theory of SCLC, similar features are attributed to a
distribution of energies of the bulk traps. Specifically, a
power-law like ($I \propto V^n$ with $n>2$) transition is
invariably attributed to a continuous distribution of trap
energies, whereas multiple discrete traps are invoked to account
for multiple transitions. Our results show that such an
interpretation is not unique and may explain inconsistencies found
in the past.

\begin{figure}[t]
\centering
\includegraphics[width=7.5cm]{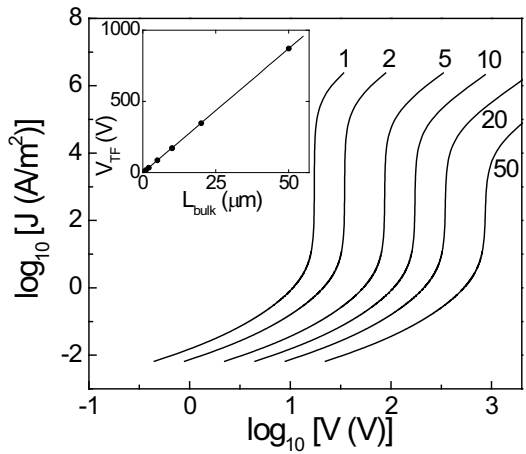}
\caption{SCLC $I-V$ curves with surface traps at the injecting
contact, for different contact separations $L$, ranging from 1 to
50 $\mu$m (as indicated in the figure). If the total number of
surface traps is larger than the total number of bulk traps, the
shape of the $I-V$ curve is essentially identical to that
predicted by the conventional SCLC theory. However, $V_{TF}$
scales linearly with $L$, as shown in the inset.
\label{simulationlength}}
\end{figure}

One more experimentally relevant finding regards the value of the
voltage $V_{TF}$ at which the transition to the trap filled limit
occurs. In the conventional theory $V_{TF} \propto N_tL^2$, which
is often used to estimate the bulk density of deep traps.
Fig.~\ref{simulationlength} shows the behavior of SCLC curves when
surface traps are present at the injecting contact. In this case,
the qualitative shape of the SCLC curves is identical to that
obtained with the conventional theory. However, the value of
$V_{TF}$ is different, and is found to scale linearly with $L$
(see inset). Thus, the experimental observation of such a linear
scaling provides a direct way to demonstrate the relevance of
surface traps.

The results discussed above can be understood in terms of the
electrostatics of the system. In very simple terms, in a SCLC
experiment a device can be thought of as a capacitor that is
charged by the applied voltage. The spatial distribution of traps
determines the capacitance of the device. Since in a parallel
plate geometry the capacitance is inversely proportional to the
distance between the charges, traps at the injecting contact have
a low capacitance (the distance between the charges corresponds to
the total thickness of the crystal $L$), so that the application
of a large voltage bias is needed to fill them. Since essentially
no charges are injected in the bulk until a sufficiently large
voltage is applied to fill all the surface traps, this results in
a strong perturbation of the electrostatics throughout the bulk of
the device, which is why surface traps at the injecting contact
have such a large influence on the $I-V$ curve. On the contrary,
surface traps close to the extracting contact have a large
capacitance, so that their filling occurs already at a very small
voltage bias. On the voltage scale used in the measurements (and
for sufficiently thick crystals) the perturbation to the
electrostatics is only minor and does not prevent the injection of
charges into the bulk. This difference directly accounts for the
asymmetry of the $I-V$ curves. It also accounts for the behavior
of $V_{TF}$ (see Fig.~\ref{simulationlength} inset), since the
capacitance of traps located at the injecting contact scales
linearly with $L$.

In conclusion, we have shown that the influence of surface traps
is an essential ingredient for the proper understanding of SCLC
experiments. The crucial point is that in high-purity samples, the
total amount of surface traps can dominate over the total amount
of bulk traps, even for sizable contact separations. Under these
conditions, charge trapped at the surface strongly modifies the
electrostatic profile inside the bulk, which in turn determines
the amplitude of the measured current. Although SCLC measurements
have been performed for over 50 years ago and their theoretical
description is now textbook material, the large effect of surface
traps on the electrostatics had never been recognized earlier.

\begin{acknowledgments}
We acknowledge FOM for financial support. The work of AFM is part
of the NWO Vernieuwingsimpuls 2000 program.
\end{acknowledgments}


\begin{thebibliography}{99}
\bibitem{parameters} In the regime that we are analyzing, the density of dopants is
much smaller than the density of traps and their energy depth is
shallower than for trap states. The conclusions of this work do
not depend on the precise values. The density of states in the
valence band is also needed in our calculations and it is taken to
be one state per molecule ($3.3 \cdot 10^{21}$ cm$^{-3}$), as
usually done in literature.
\end{thebibliography}
\end{document}